# Role of substrate surface morphology on the performance of graphene inks for flexible electronics

Jasper Ruhkopf[1,2], Simon Sawallich[2,3], Michael Nagel[3], Martin Otto[1], Ulrich Plachetka[1], Tom Kremers[4], Uwe Schnakenberg[4], Satender Kataria[2,*], Max C. Lemme[1,2,*]

[1]AMO GmbH, Advanced Microelectronics Center Aachen (AMICA), Otto-Blumenthal-Str. 25. 52074 Aachen, Germany
[2]RWTH Aachen University, Chair of Electronic Devices, Otto-Blumenthal-Str. 2 52074 Aachen, Germany
[3]Protemics GmbH, Otto-Blumenthal-Str. 25. 52074 Aachen, Germany
[4]RWTH Aachen University, Institute of Materials in Electrical Engineering 1, Sommerfeldstr. 24, 52074 Aachen, Germany
*Corresponding author: Satender.kataria@eld.rwth-aachen.de, lemme@amo.de

**Abstract**

Two-dimensional (2D) materials like graphene are seen as potential candidates for fabricating electronic devices and circuits on flexible substrates. Inks or dispersions of 2D materials can be deposited on flexible substrates by large-scale coating techniques like inkjet printing and spray coating. One of the main issues in coating processes is non-uniform deposition of inks which may lead to large variations of properties across the substrates. Here, we investigate the role of surface morphology on the performance of graphene ink deposited on different paper substrates with specific top coatings. Substrates with good wetting properties result in reproducible thin films and electrical properties with low sheet resistance. The correct choice of surface morphology enables high-performance films without post-deposition annealing or treatment. Scanning Terahertz time-domain spectroscopy (THz-TDS) is introduced to evaluate both the uniformity and the local conductivity of graphene inks on paper. A paper-based strain gauge is demonstrated and a variable resistor acts as an on-off switch for operating an LED. Customized surfaces can thus help in unleashing the full potential of ink-based 2D materials.

**Introduction**

Two dimensional (2D) layered materials such as graphene have received significant attention and are considered for applications in a broad spectrum of fields spanning from electronics to biology [1,2]. 2D materials possess a wide range of interesting properties, which include high electrical and thermal conductivity, strong light-matter interaction for photodetection and light emission and high mechanical strength and flexibility, making them a potential candidate for flexible devices and wearables [3]. 2D materials can be obtained by different methods such as mechanical exfoliation, epitaxial growth, chemical vapor deposition techniques and liquid phase exfoliation [4]. Intense research on the latter method of production has led to inks and dispersions of 2D materials, which are now commercially available in large quantities [5–7]. Such liquid dispersions have been proposed for printed electronics [8–11] and as composites for energy [12,13] and health applications [14,15]. There exist numerous reports on the deposition of 2D inks using techniques like inkjet printing, spray coating and drop-casting [16–18]. Despite great progress, 2D-material-based inks still face limits regarding uniform deposition and consistent electrical performance of solvent based inks [19]. For applications as flexible devices, 2D inks are generally required to be deposited on flexible substrates which are either polymer based such as Kapton, Polydimethylsiloxane (PDMS) and Polyethylene terephthalate (PET) or paper based. Generally, such substrates have low surface energy [20,21] which leads to non-uniform deposition of inks and thus gives rise to issues such as coffee-ring effects [22,23] or incomplete coalescence of single droplets [24]. Often, additional processes like thermal or laser annealing are required for improving the properties of the as-deposited inks to obtain reliable and reproducible [16,25,26]. Therefore, a basic understanding and tailoring of the substrate surface is essential for exploiting the full potential of 2D inks in flexible and wearable electronics.

Here, we demonstrate how surface morphology of the substrate plays a significant and crucial role on uniform deposition of graphene inks on flexible paper substrates. We find that a surface with good wetting properties and porous microstructure leads to uniform deposition of inks with desirable electrical properties without requiring further post-deposition treatments. Raman spectroscopy and THz-TDS confirm the uniform deposition and, more importantly, reproducible properties of the inks. Finally, the proposed method is applied to graphene ink on paper based flexible strain sensors and variable resistors, which can act as on-off switches in a simple circuit.

**Experimental**

We used commercial graphene dispersion (Thomas Swan Elicarb) for the experiments. The dispersion contains few layer graphene flakes dispersed in water and stabilized with sodium cholate as a surfactant. The flake thickness is five to eight layers in average while the mean lateral dimension is around 1 µm. The concentration of the dispersion is 1 g/L. A detailed description of the ink production method can be found in [5]. We have used electronic grade paper (Felix Schoeller) p_e:smart type 1, 2 and 3 as flexible substrates. The material has a paper core with a smoothening surface coating on the front- and backside. A thin top coating layer, dense for paper type 1 and nano-porous for type 2 and 3, is present as well. According to the paper type, the samples were named sample S1 for paper type 1, S2 for type 2 and S3 for type 3 paper. An additional letter e.g. sample S2a, S2b indicates samples fabricated on paper type 2 in different runs using the same graphene ink.

The samples (20x20 mm²) were coated by spray deposition using a Badger 200 NH airbrush gun with 0.76 mm nozzle diameter and compressed air as the propellant gas. In order to achieve a uniform layer, the spray parameters were optimized to 2.5 bar backpressure and a working distance of 10-15 cm. The ink flow rate was kept at a constant low level during the whole coating process. An average of 0.25 ml ink per cm² substrate was used through all the experiments, to ensure that the same amount of material was deposited on all samples. After the deposition process, the samples were dried under ambient conditions for 30 minutes. No further post-treatment such as high temperature or laser annealing was conducted. The sample surface morphology was characterized using atomic force microscope (AFM) (Veeco Instruments) before and after ink deposition. The samples were then cut into 10x10 mm² pieces and fixed on a 20x20 mm² silicon chip with double-sided tape to avoid measurement errors caused by a curved sample. Raman spectroscopy was carried out using a WITEC 300R system and confirmed the uniformity of the deposited inks. A 532 nm wavelength laser with a power of 1 mW was employed for the Raman measurements. For the estimation of graphene layer thicknesses, a ZEISS Supra 60VP FE-SEM was utilized to take cross-section images of the samples.

The sheet resistance of as-coated samples was measured using the four-point probe (4PP) technique (CMT-SR2000N, Advanced Instrument Technology). A Jandel probe head equipped with Tungsten Carbide needles (tip radius of 100 µm and 1 mm needle spacing) was mounted. Several measurements were taken at different positions close to the center. The impressed current was set to 1 mA. A correction factor of 4.53 was applied for the sheet resistance calculations. The wetting abilities of the paper substrates were measured using contact angle measurements. The contact angle of ink droplets on different substrates was measured with a home-built goniometer setup consisting of a moveable stage, a camera and a housing. A droplet of 5.5 µL was casted on various substrates with an Eppendorf pipette. The apparent contact angle was calculated using a customized image analyzing software evaluating the droplet image taken with the camera. The evolution of the contact angle over time was also studied by capturing the droplet image successively for 3 minutes at 30 second intervals.

Electrical characterization of the strained/unstrained samples was carried out with a Keithley semiconductor analyzer SC4200. Compressive and tensile strain was applied by fixing the samples to a stainless-steel bending beam and loading the beam with different weights. Terahertz (THz) conductivity measurements were performed in a transmission setup based on a pump/probe scheme, using laser pulses with 100 fs duration at a central wavelength of 780 nm and a bias-free bimetal grating Schottky-field-emitter for THz generation. Sub-wavelength spatial resolution is achieved by using a near-field microprobe (TeraSpike TD-800-X-HRS, Protemics GmbH) for photo-conductive THz detection, which was scanned across the investigated coated samples in a few µm distance. The sheet resistance $R_S$ of the conductive layer on top of the paper substrate was calculated from the measured THz transmission amplitude reduction generated by the coating layer using the Tinkham equation [27]. The direct substrate transmission value was measured through a bare paper substrate alone.

**Results and discussion**

Figs. 1a-1c show a comparison of the surface morphologies of the pristine paper samples obtained through AFM by scanning an area of 5 µm × 5 µm. The surface of uncoated

paper 1 (sample S1) is smooth, while paper 2 (sample S2) and paper 3 (sample S3) exhibit similar morphologies with small nano-pores in the range of ~200 nm diameter. The measured root mean square (rms) of S1 was 2.17 nm, which is a factor of five lower compared to that of sample S2 (9.74 nm) and S3 (11.71 nm). The AFM scans were repeated after spray coating the three different papers with graphene ink (Fig. 1d-1f). The surface morphology of graphene coated papers has been found to be almost identical on all three substrates with an rms roughness in the range of ~60 nm. The increase in roughness is mainly attributed to the random orientation of the few-layer graphene flakes, which are partly aligned flat on the substrate surface but were also found standing upright. Furthermore, separate few-layer graphene flakes with lateral dimensions of ~1 µm can be seen in the images (Fig. 1d-1f).

The microstructure of the paper coating layers and thereon deposited graphene flakes was studied by SEM. Cross-section images of sample S1-S3 were taken at similar levels of magnification, to compare the layer stacks and estimate the graphene layer thickness (Figs. 2a-2c). The images reveal, that for roughly the same given amount of material deposited, the graphene layer on sample S1 is thicker (~2 µm) compared to S2 and S3 (<1 µm). Furthermore, the structural difference of porous (S2 and S3) and non-porous (S1) paper coatings can be observed. The porous coatings can be identified by vertical grooves in the cross-sectional area, while their absence characterizes the non-porous layer. In Fig. 2d, the microstructure of graphene flakes deposited on sample S2 is shown at higher magnification. A densely packed stack of individual graphene flakes is visible.

The properties of the graphene ink on the paper substrates were further investigated through Raman spectroscopy. An area of 40 µm x 40 µm (marked by the red rectangle in the optical micrograph shown in Fig. 3a, sample S2) was scanned with a step size of 1 μm. Fig. 3b shows a point spectrum obtained at the center of the area to verify the presence of few-layer graphene. A 2D/G intensity ratio of less than one, with sharp G and 2D peaks confirmed that the flakes are indeed crystalline multilayer graphene [28]. The presence of a strong D peak is typical for edge defects in graphene dispersions [29,30]. The disorder related D' peak is visible as well [31]. Fig. 3c shows the map of the 2D/G intensity ratio extracted from the Raman scan. The uniform intensity ratio (0.53 to 0.55) over the whole scanned area confirms the structural uniformity of the deposited inks. Fig. 3d shows a map of the D/G ratio and it is predominantly around 0.3 across the sample.

The electrical properties of the deposited graphene inks were characterized to assess the suitability of different papers as substrates for electronic applications. The sheet resistance of the as-deposited inks was measured using a four-point probe setup. Table 1 summarizes the sheet resistance values for the different paper substrates. Sample S1 exhibits a high sheet resistance of 700 kΩ/sq on average. On the other hand, sample S2 and S3 have much lower sheet resistances of approximately 3 kΩ/sq. This is in stark contrast to the similar surface morphology observed in all three samples through AFM and Raman measurements: This suggests that the substrate surface has a significant impact on the electrical performance of graphene inks regardless of the observed uniform surface topography. From an application point of view, the sheet resistance of sample S1 is much too high, while the samples S2 and S3 exhibit competitive values to recently reported data [32]. It should be highlighted that no post-deposition treatment was conducted prior to the electrical measurements in our current experiments, whereas such treatments have been widely used and shown to greatly improve electrical properties of graphene inks. The porous coating of paper type 2 and type 3 thus enables application relevant conductivities of graphene inks utilizing low-temperature processing. This is a fundamental achievement, as porous substrates will allow low-cost roll-to-roll production of flexible substrates with conductive graphene coatings regardless of the ink deposition technique used.

**THz characterization:** THz-TDS sheet resistance mapping was carried out on paper type 2 to evaluate the spatial homogeneity of the electrical properties of the deposited ink (sample S2a). A simplified drawing of the THz-TDS measurement setup is provided in Fig. 4a. The mapping yielded an average sheet resistance of 258 $\Omega$/sq. (Fig. 4b). The mapping shows homogeneous coating of the substrate, which supports the previous observations of morphology and structural homogeneity made through AFM and Raman measurements. The statistical distribution of the measured sheet resistance is shown in Fig. 4c with a standard deviation $\sigma$ of 18 $\Omega$/sq. 4PP sheet resistance measurements done on S2a (Table 2), resulted in an average value of 1.54 k$\Omega$/sq. The value differs from the sheet resistance value of 3.07 k$\Omega$/sq measured on sample S2 fabricated on the same substrate in a different batch. This is attributed to the manual application method of the coating with a handheld airbrush. The considerable discrepancy (around a factor of 6) of the sheet resistance values of S2a recorded by two different techniques can be explained using the employed extraction method. By measuring the THz-absorption by free charge carriers in the conductive layer of interest (i.e. the graphene coating in this case), the conductivity/sheet resistance of the intrinsic graphene is obtained [33]. Though the results are low-frequency resistance or quasi-DC values, it is important to note that the excited charge carriers will only move distances of up to 100 nm. As a result, there is no net induced DC-current flow in the graphene flake film. On the other hand, current flows between the two outer needles separated by a distance of 3 mm during the 4PP measurements. This means that the conductivity extracted through THz TDS is a local conductivity in the sub-µm-scale (i.e. often within a flake), while the 4PP measurement averages the current flow through an area much larger than the average flake size. Thus, charge carriers need to pass many high resistance flake-to-flake-junctions to travel between the probes and hence a higher sheet resistance is measured by 4PP.

**Contact angle measurements:** The drying mechanism for inks on different surfaces was investigated through contact angle measurements. The evolution of the contact angles over time was recorded for pristine paper type 1, paper 2 and paper 3 (Fig. 5a-c). Droplets of 5.5 µl graphene ink were deposited on the substrates with an Eppendorf pipette. The ink droplet on paper 2 had an initial contact angle of 63.6 °, which was considerably lower than on paper 1 (81.8 °). Paper 3 even shows a lower contact angle (39.8 °). This result demonstrates the higher hydrophilicity of the porous paper top coating. During the first 180 s of drying time, the droplet's contact angle decreased while the droplet diameter was constant due to contact line pinning [22]. The contact angle decreased much faster on paper 2 and 3, which can be explained by a vertical flow of solvent into the porous top-coating. The proposed drying mechanism, in case of porous coating, is provided as a schematic in Fig. 5d: The solvent is drained by capillary forces created by the nanometer sized pores, which takes place simultaneously with the solvent evaporation process though the latter is happening much slower[34]. During airbrush coating, pL droplets are deposited on the substrate surface as measured in Scardaci et al.[35]. For small droplets in the pL range, the timescale of both drying mechanism was investigated by Tan[34]. The author found that solvent absorption by the porous substrate is completed within milliseconds while the evaporation of residual solvent happens in the seconds range. However, for µL droplets as used for the contact angle measurements, the time scale of solvent absorption shifts towards the seconds range, because the absorption time is proportional to a power law dependence of the initial droplet diameter[34]. The observed wetting and liquid penetration properties are in agreement with recent studies of porous paper based materials [36,37]. On paper 1, the solvent cannot enter the substrate, therefore, evaporation is the dominating mechanism involved in droplet drying. Wu et al. found a similar effect for thermoplastic electrically conductive adhesives deposited on various paper substrates with different grades of porosity [38]. On all three papers, the decrease of contact angle was found to be linear for a given timeframe. It took approximately 30 min for all droplets to dry completely, observable by naked eye. On paper type 1, the coffee-ring effect is clearly visible, with a thicker flake layer at the droplet perimeter compared to the droplet center. The dried ink droplet on paper 2 is homogeneous and no coffee-ring formation is observed. The faster drying of the ink on paper 3 leads to the generation of "cobweb"-like structures. The occurrence of this effect seems to depend on the droplet size, as it was not present for smaller droplets. As studied by Pack et al., the pore size determines whether a coffee-ring is formed or uniform deposition could be observed[39]. A pore size of 200 nm, which was measured by AFM on sample S2 and S3, generates a uniform layer which indeed could be validated experimentally.

This makes the porous paper 2 and 3 the most suitable substrates for graphene ink deposition of those studied in this article.

We assume that the drying mechanisms described above directly affect the microstructure of the deposited graphene flake layer. Capillary forces, present during ink deposition on porous substrates generate a well-aligned, densely packed graphene flake layer. In absence of these forces, the graphene flakes are randomly oriented and exhibit a high porosity which was observed on non-porous substrates. The density of the graphene layers also explains the difference in sheet resistance. This is similar to a study by Huang et al., which reports the sheet resistance of graphene flake layers after compressing the layers by applying different amounts of pressure[40]. The higher the flake density of the layers, the lower was the measured sheet resistance. It is evident from the comparison of the THz-TDS and the 4PP measurements that the overall resistance of the layers is dominated by flake-to-flake-junctions, which form a conductive percolation network [41]. It has been shown that electron tunneling is the predominant mechanism for the electrical connection of two neighboring graphene flakes [42]. The general equation for electron tunneling, derived by Simmons [43], states that the tunnel resistivity $\rho$ is proportional to $e^d$ with $d$ defined as the flake-flake distance. A densely packed layer has a lower average flake-flake distance compared to a porous layer and thus shows a lower sheet resistance.

Bending beam measurements on sample S2b (paper type 2 substrate) were carried out to study its performance as strain gauge. A significant decrease of resistance for compressive strain (inward bending) and an increase in resistance for tensile strain (outward stretching) can be clearly observed. Fig. 6a shows the relative change in resistance ΔR/$R_0$ for several load/unload cycles of the bending beam, where R is the resistance under strained condition and $R_0$ is the resistance of the unstrained sample.

Tensile strain was applied by loading the beam with a weight of 2 kg which equals a strain of approximately ε = 4.25e-4 for the given parameters. A gauge factor (GF) of 8 was extracted, using the equation $GF = \frac{(R-R_0)/R_0}{\varepsilon}$. As shown by Hempel et al. [44], strain induces movement of the flakes, i.e. the overlapping areas and distances between connected flakes change considerably. The compressive strain decreases the flake-flake distance $d$ and therefore reduces the tunneling resistivity according to the Simmons-equation, which is in line with our observations as well. Utilizing the bending direction dependent induced changes in resistance, we have applied the graphene coated paper (sample S2) as a resistor switch. A 9 V battery was connected to the clamp-contacted paper in series with an LED. By bending the paper the LED can be switched on and off (Fig. 6b and 6c, respectively). This concept was generally shown by Casiraghi et al., without discussing the influence of the paper substrate's morphology on the electrical properties of the deposited graphene flake layer [45].

**Conclusion:**

We have studied the influence of substrate surface morphology on the structural and electrical properties of graphene inks deposited thereon. Commercial graphene ink was spray coated onto different paper substrates with porous and non-porous surface coating. The topography of the graphene ink layers was characterized by AFM and showed uniform deposition of the graphene flakes, independent of the substrate. Raman mapping confirmed the graphitic nature of the graphene flakes and visualized the layer homogeneity. The papers with porous coating showed superior characteristics as substrates for graphene flake layers with considerably lower sheet resistance values, measured by the 4-point-probe method and THz time-domain spectroscopy. Capillary forces present in the porous coatings drain the ink's solvent uniformly, resulting in densely packed graphene layers. Thus low sheet resistance without any post-deposition thermal annealing steps was observed. This hypothesis is further supported by contact angle measurements performed during the ink drying process. A paper-based strain gauge was fabricated from the porous substrate, utilizing the percolation-type charge transport in the deposited graphene ink. This effect was also deployed as variable resistor in a small electronic circuit. The study demonstrates the importance of nanoscale surface morphology of flexible substrates to exploit the inherent properties of inks and dispersions of 2D materials.

**Acknowledgements**

The authors thank Felix Schoeller company for providing the paper samples. Funding through the European Commission (Graphene Flagship, 785219), the German Ministry for Economic Affairs and Energy BMWi (PROTONLY, 49VF170038) and the German Ministry for Education and Research (GIMMIK, 03XP0210) is gratefully acknowledged.

**Figures**

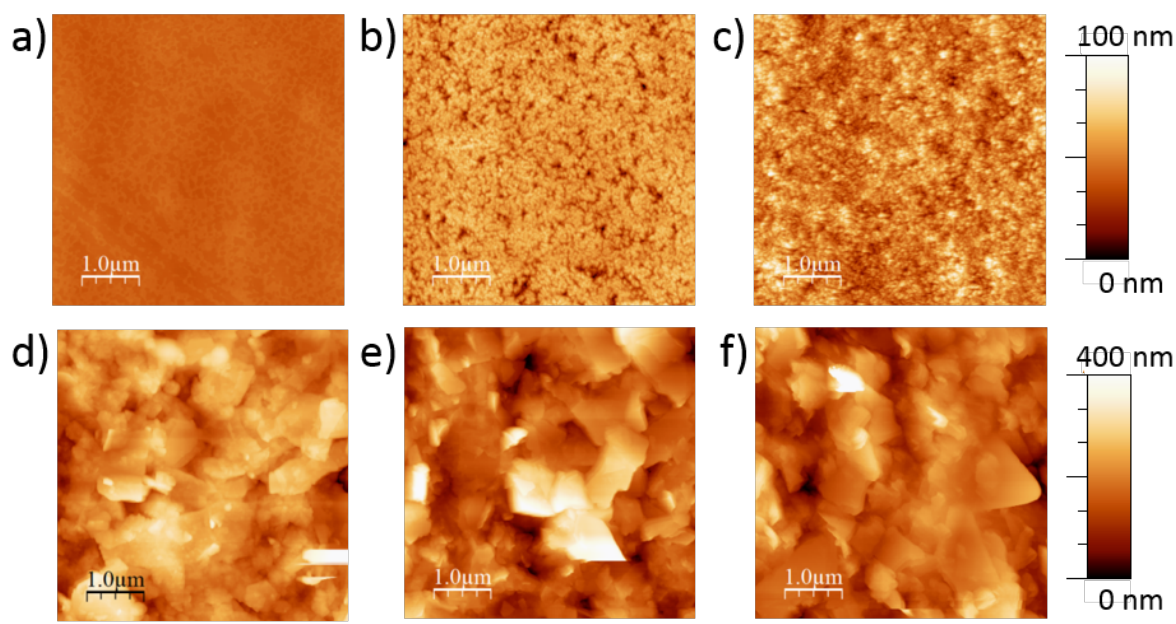

**Figure 1.** AFM scans of uncoated and graphene ink coated paper substrate. (a)-(c) Scan of S1-S3 uncoated. S1 shows a dense smooth surface while S2 and S3 contain nano-pores of approximately 200 nm diameter. (d)-(f) Graphene ink coated paper samples S1-S3 with similar morphology. The lateral dimension of graphene flakes is ~1.0 µm.

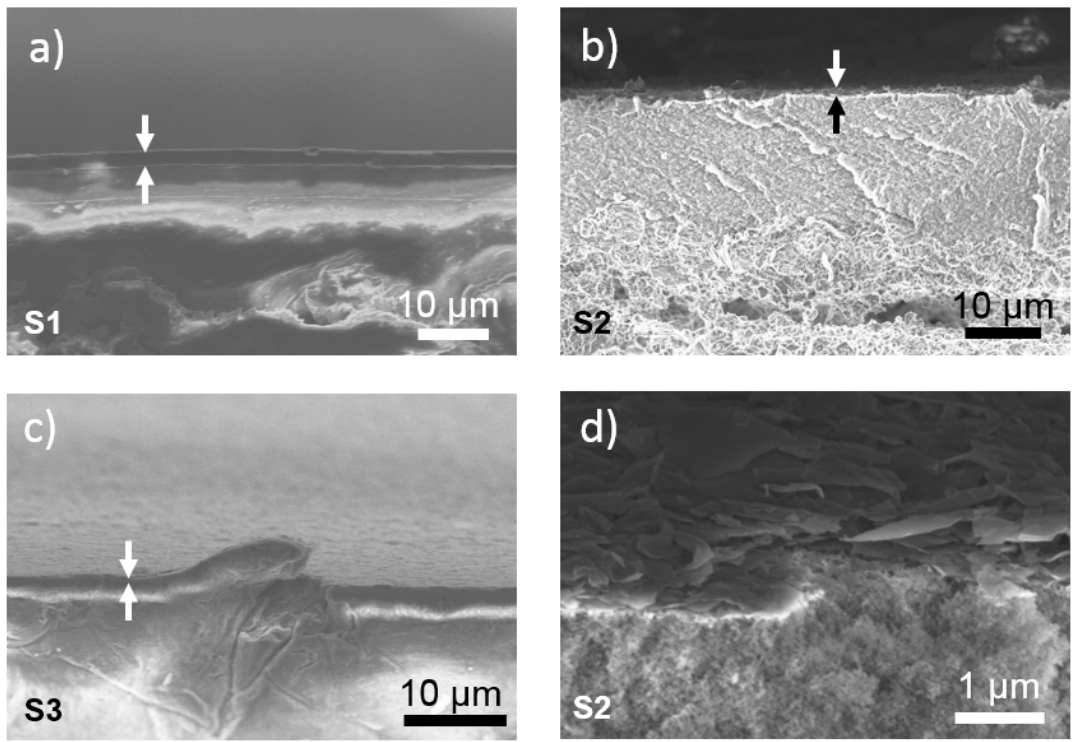

**Figure 2.** SEM cross-section images of graphene ink coated papers S1-S3. (a)-(c) Images of S1-S3 were taken at ~5kx magnification, to compare the layer stack and determine the graphene thickness. The graphene layers are indicated by arrows. In (a), the graphene thickness is in the range of 2 µm while in (b) and (c) the thickness is below 1 µm (d) Image of S2 taken at 50kx magnification. The densely packed graphene flakes can be observed.

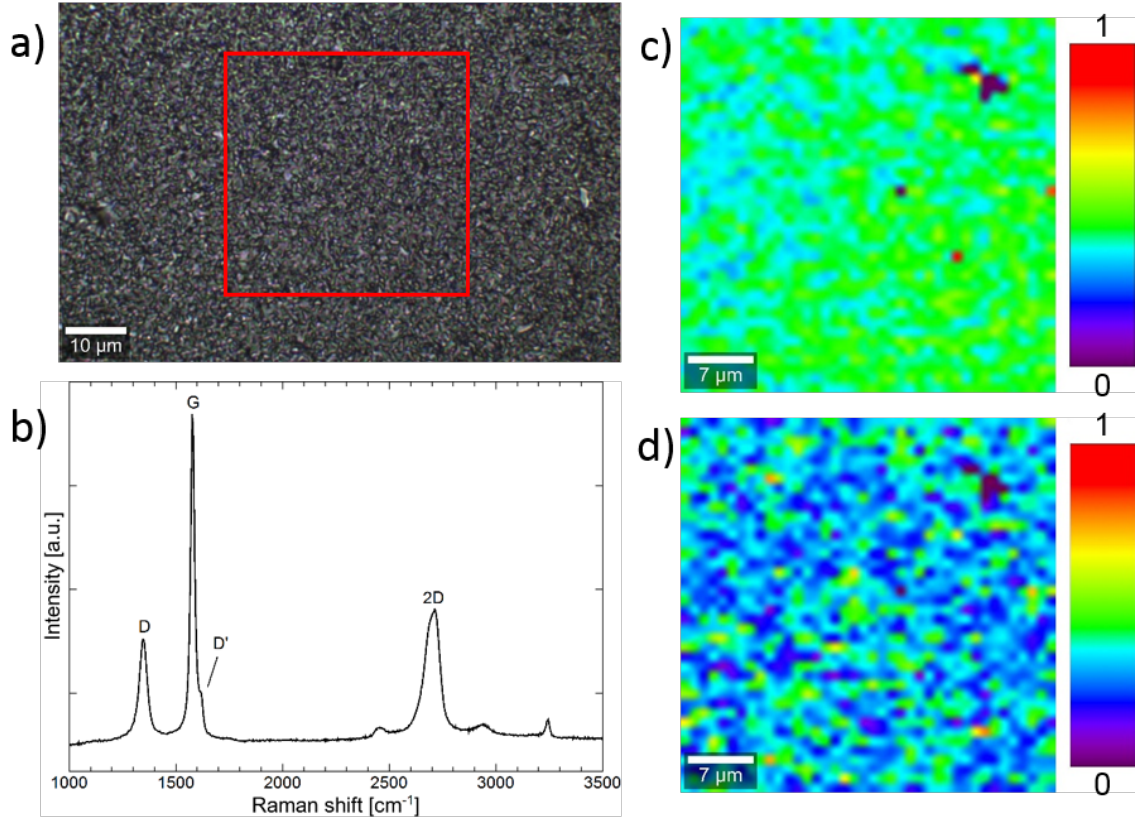

**Figure 3.** Raman characterization of graphene coated sample S2. (a) Optical micrograph. Red rectangle shows analyzed sample area of 40 μm x 40 μm. (b) Average point Raman spectrum of the graphene ink. (c) Raman map of 2D/G intensity ratio shows homogeneous flake structure over scanned area. (d) Raman map of D/G intensity ratio.

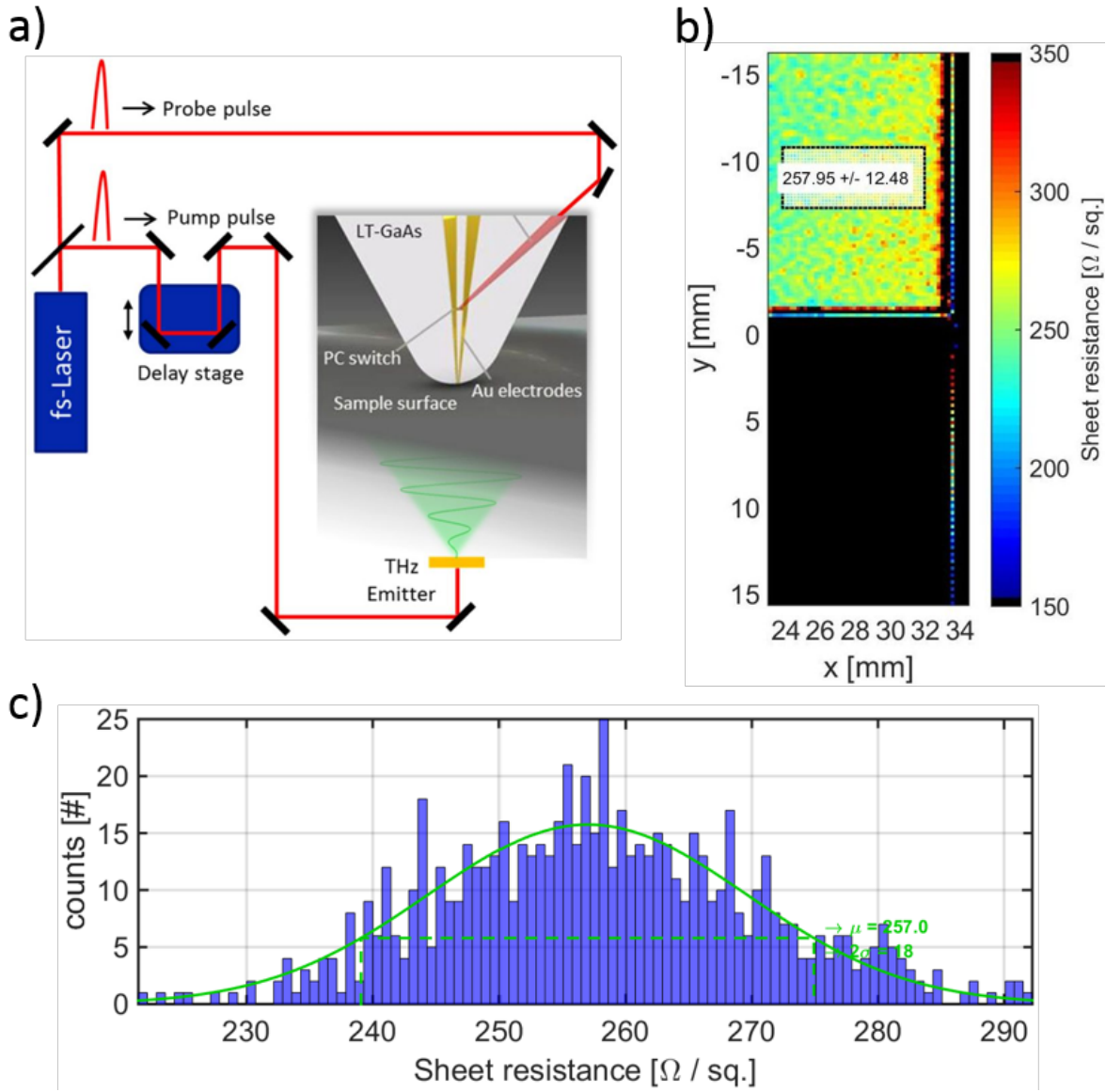

**Figure 4.** Sheet resistance mapping by THz spectroscopy. (a) Measurement setup. (b) Sheet resistance map. The top sample is ink coated paper S2a (avg. sheet resistance 258 Ω/□), the bottom one uncoated. (c) Statistical distribution of sheet resistance on S2a.

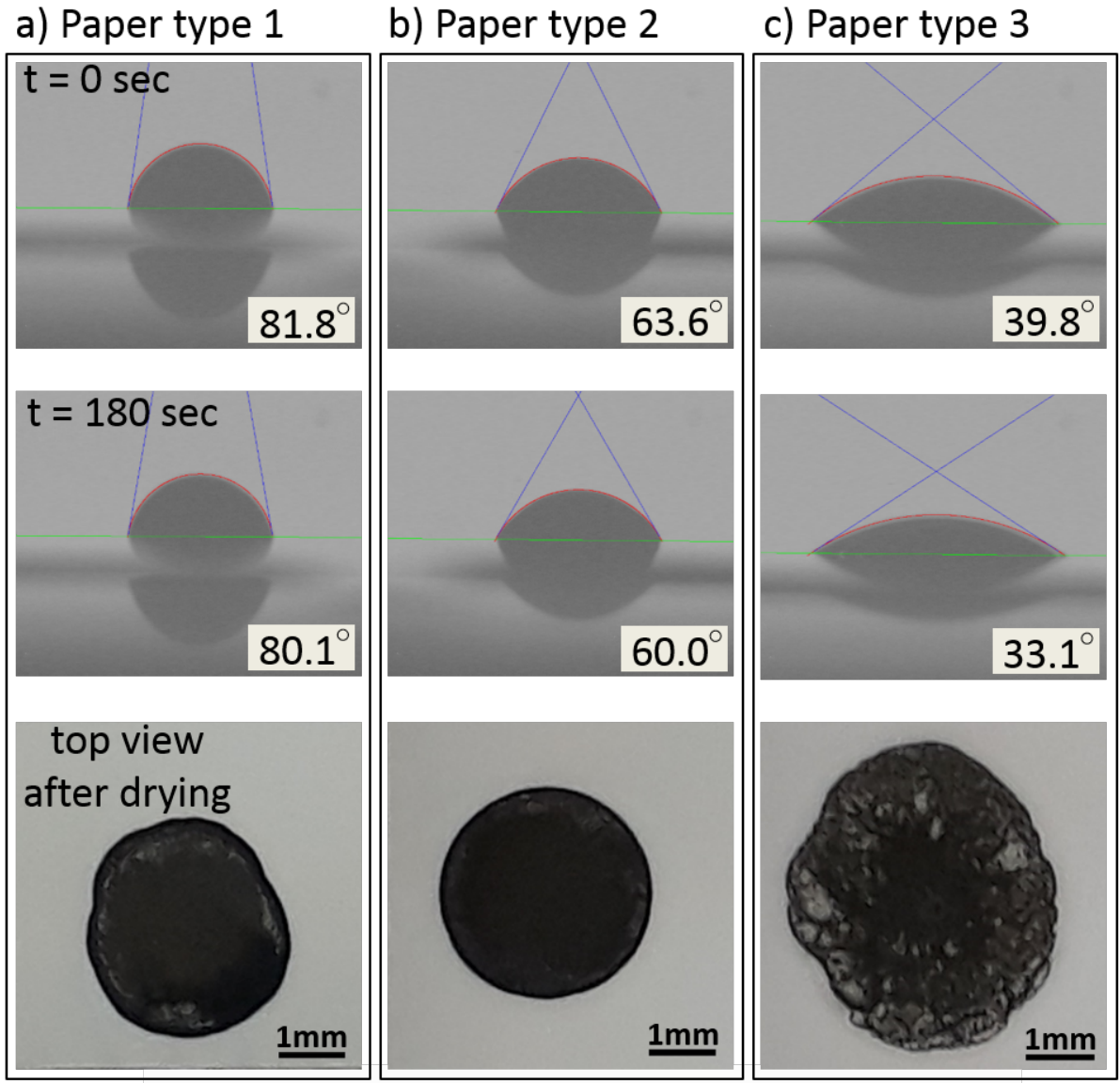

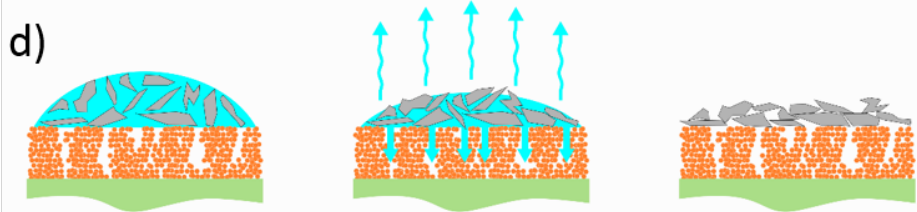

**Figure 5.** Time dependent contact angle measurements after droplet deposition and after 3 min drying together with top view photographs after complete droplet drying. (a) Paper type 1 substrate (b) Paper type 2 (c) Paper type 3. Paper type 2 facilitates homogeneous droplet drying, while a coffee-ring effect is clearly visible on Paper type 1 and 3. (d) Proposed drying mechanism of graphene ink on porous surfaces. Water can enter the substrate's pores for improved drying condition.

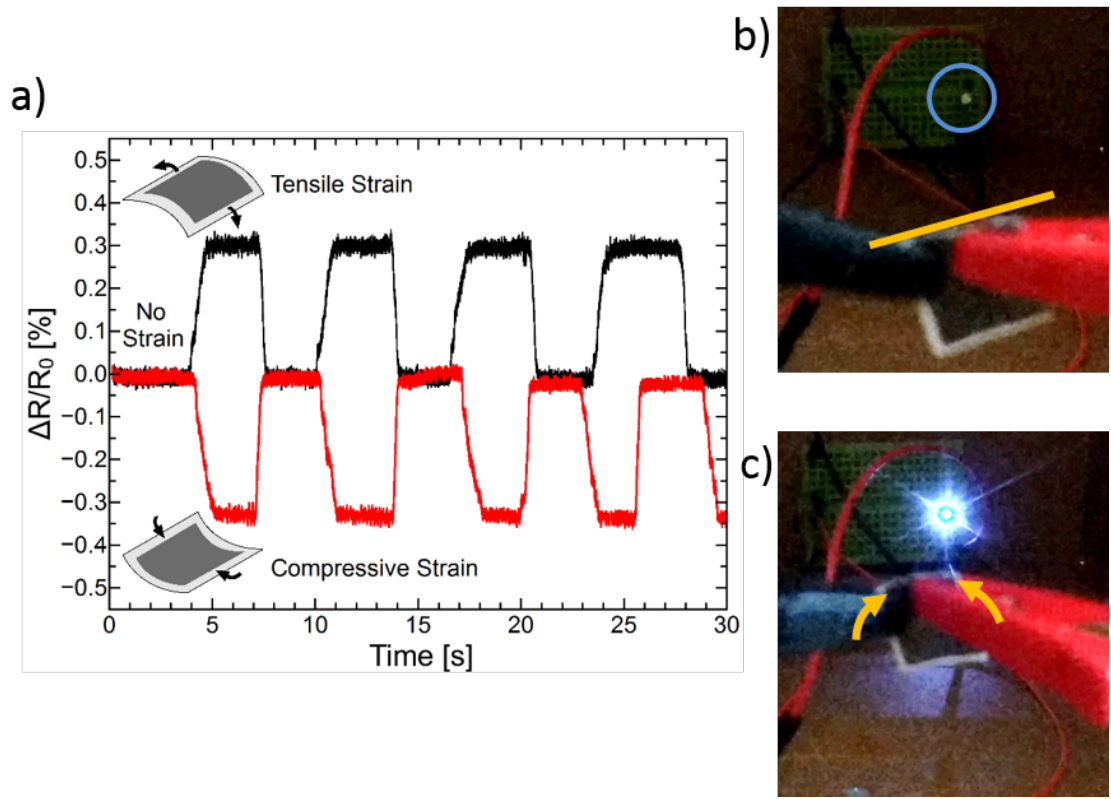

**Figure 6.** Bending beam measurement of paper sample S2b and switching application (sample S2). (a) Electrical characterization of S2b with tensile (outward bending) and compressive strain (inward bending) applied. The percolation path of graphene flakes network changes by bending and thus the sample resistance. (b) and (c) show the application as switch. In (b) the switch is in off-state (sample kept straight) in (c) in on-state (sample inward bended).

**Tables**

**Table 1.** Sheet resistance mapping by four-point-probe method. Comparison of ink coated samples S1-S3.

| Sample | Sheet resistance [kΩ/□] | | | |
|---|---|---|---|---|
| | Position 1 | Position 2 | Position 3 | Average |
| Sample S1 | 671.50 | 765.90 | 687.00 | 708.10 |
| Sample S2 | 3.12 | 2.83 | 3.27 | 3.07 |
| Sample S3 | 3.30 | 3.34 | 3.24 | 3.29 |

**Table 2.** Sheet resistance mapping by four-point-probe method of sample S2a.

| Measurement position | Sheet resistance [kΩ/□] |
|---|---|
| 1 | 1.62 |
| 2 | 1.59 |
| 3 | 1.75 |
| 4 | 1.34 |
| 5 | 1.49 |
| 6 | 1.55 |
| 7 | 1.37 |
| 8 | 1.51 |
| 9 | 1.63 |

**For Table of Contents Only**

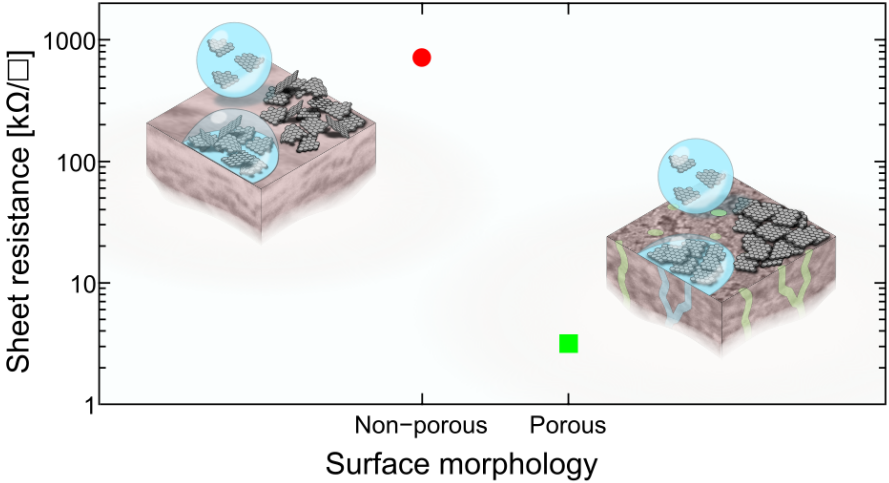

**TOC graphic.**

27